 \newcommand{\be}{\begin{equation}}\newcommand{\ee}{\end{equation}}
\newcommand{\bea}{\begin{eqnarray}}\newcommand{\eea}{\end{eqnarray}}
\newcommand{\nn}{\nonumber}\newcommand{\p}[1]{(\ref{#1})}
\newcommand{\lb}{\label}

\newcommand{\cA}{{\cal A}}
\newcommand{\bcA}{\bar{\cal A}}
\newcommand{\cW}{{\cal W}}
\newcommand{\bcW}{\bar{\cal W}}

\newcommand\T{\mbox{Tr}\;}

\newcommand\sY{{\s Y}}
\newcommand\s{\scriptscriptstyle}
\newcommand\q{\quad}
\newcommand\qq{\qquad}

\newcommand{\pp}{{=\!\!\!|}}
\newcommand{\xp}{x^\pp}
\newcommand{\xm}{x^=}

 \newcommand{\PY}{\partial_\sY}
 \newcommand{\bPY}{\bar\partial_\sY}
 \newcommand{\Pp}{\partial_\pp}
\newcommand{\Pm}{\partial_=}

\newcommand{\tpi}{\theta^+_i}

\newcommand{\tmi}{\theta^-_i}

\newcommand{\btpi}{\bar{\theta}^{i+}}

\newcommand{\btmi}{\bar{\theta}^{i-}}

\newcommand{\bppo}{\bar{\partial}_{1+}}
\newcommand{\bppt}{\bar{\partial}_{2+}}
\newcommand{\bpmo}{\bar{\partial}_{1-}}
\newcommand{\bpmt}{\bar{\partial}_{2-}}
\newcommand{\bpmh}{\bar{\partial}_{3-}}
\newcommand{\bpph}{\bar{\partial}_{3+}}
\newcommand{\ppo}{\partial^1_+}
\newcommand{\ppt}{\partial^2_+}
\newcommand{\pmo}{\partial^1_-}
\newcommand{\pmt}{\partial^2_-}
\newcommand{\pmh}{\partial^3_-}
\newcommand{\pph}{\partial^3_+}

\newcommand{\Dpi}{D_+^i}
\newcommand{\Dpk}{D_+^k}
\newcommand{\Dpl}{D_+^l}

\newcommand{\Dmk}{D_-^k}
\newcommand{\Dml}{D_-^l}
\newcommand{\bDpi}{\bar{D}_{i+}}
\newcommand{\bDpk}{\bar{D}_{k+}}
\newcommand{\bDpl}{\bar{D}_{l+}}

\newcommand{\bDmk}{\bar{D}_{k-}}
\newcommand{\bDml}{\bar{D}_{l-}}

\newcommand{\tpo}{\theta^+_1}
\newcommand{\tpt}{\theta^+_2}
\newcommand{\tmo}{\theta^-_1}
\newcommand{\tmt}{\theta^-_2}
\newcommand{\tmh}{\theta^-_3}
\newcommand{\tph}{\theta^+_3}
\newcommand{\btpo}{\bar\theta^{1+}}
\newcommand{\btmo}{\bar\theta^{1-}}
\newcommand{\btpt}{\bar\theta^{2+}}
\newcommand{\btmt}{\bar\theta^{2-}}
\newcommand{\btph}{\bar\theta^{3+}}
\newcommand{\btmh}{\bar\theta^{3-}}
\newcommand{\Dpo}{D_+^1}
\newcommand{\Dmo}{D_-^1}
\newcommand{\Dpt}{D_+^2}

\newcommand{\bDph}{\bar{D}_{3+}}
\newcommand{\bDmh}{\bar{D}_{3-}}

\newcommand{\bDpt}{\bar{D}_{2+}}

\newcommand{\Dot}{D^1_2}
\newcommand{\Dto}{D^2_1}
\newcommand{\Doh}{D^1_3}
\newcommand{\Dho}{D^3_1}
\newcommand{\Dth}{D^2_3}
\newcommand{\Dht}{D^3_2}
\newcommand{\hDot}{\hat{D}^1_2}

\newcommand{\hDoh}{\hat{D}^1_3}

\newcommand{\hDth}{\hat{D}^2_3}

\documentstyle[12pt]{article}

\def\theequation{\arabic{section}.\arabic{equation}}
\begin{document}
\begin{center}
{\large\bf
HARMONIC-SUPERSPACE METHOD OF  SOLVING  N=3 SUPER-YANG-MILLS EQUATIONS   }
\vspace{0.5cm}

{\bf Ji\v{r}\'{\i} Niederle~$^a$\footnote{E-mail: niederle@fzu.cz}
 and Boris Zupnik~$^b$\footnote{E-mail: zupnik@thsun1.jinr.ru}}\\

{\it $^a$Institute of Physics, Academy of Sciences of the Czech Republic,
Prague 8, CZ 182 21, Czech Republic}\\
{\it$^b$ Bogoliubov Laboratory of Theoretical Physics, Joint Institute
for Nuclear Research, Dubna, Moscow Region, 141980, Russia}\\
\end{center}
\begin{abstract}
We analyze the superfield constraints of the $D{=}4,~N{=}3$ SYM-theory
using light-cone gauge conditions. The $SU(3)/U(1)\times U(1)$ harmonic
variables are interpreted as auxiliary spectral parameters, and the
transform to the  harmonic-superspace representation is considered. The
harmonic superfield equations of motion are drastically simplified in our
gauge, in particular, the basic matrix of the harmonic transform and
the corresponding harmonic analytic gauge connections become
nilpotent on-shell. It is shown that these harmonic SYM-equations
are equivalent to the finite set of solvable linear iterative equations.
\end{abstract}



\renewcommand{\thefootnote}{\arabic{footnote}}
\setcounter{footnote}0
\setcounter{equation}0
\section{Introduction}

It is well known that the geometric superfield constraints in the
$D=4, N=3$ super-Yang-Mills theory are equivalent to the equations of
motion of the corresponding component fields \cite{So}. The special
projections of these constraints can be interpreted as conditions for
zero curvatures associated with the Grassmann  covariant derivatives
\cite{Wi,GIKOS,RS}.

The possible connection of these zero-curvature conditions with
integrability or solvability of the super-Yang-Mills (SYM) theories with
16 or 12 supercharges  has been discussed more than 20 years in the
framework of different superfield approaches (see, e.g.
\cite{Wi}-\cite{DO1}). Here we shall use the harmonic-superspace method
\cite{GIKOS} to analyze the solutions of the $N=3$ SYM-equations. (The
preliminary short version of this work appears as Ref.\cite{NZ}.)

The harmonic-superspace (HSS) method has been introduced first to
solve the $D=4,~N=2$ off-shell superfield constraints \cite{GIK1}.
Harmonic variables are analogous to  twistor or auxiliary spectral
variables used in integrable models. Harmonic and twistor methods give
the explicit constructions of the general solutions to zero-curvature
conditions in terms of independent functions  on special {\it analytic}
(super)spaces satisfying the generalized Cauchy-Riemann  analyticity
conditions. For instance, the off-shell $N=2$ superfield constraints
have been solved via the conditions of the Grassmann (G-) analyticity
\cite{GIK1}.

In the standard harmonic formulation of the  $D=4, N=2$ SYM-theory,
the basic harmonic connection is G-analytic  \cite{GIK1}, and the 2-nd
one ( via the harmonic zero-curvature condition) appears to be a
nonlinear function of the basic connection. The $N=2$ equation of motion
is linearly dependent on the 2-nd harmonic connection, but it is the
nonlinear equation for the basic connection \cite{Zu2}. It has been shown
in Ref.\cite{Zu4} that one can alternatively choose the 2-nd harmonic
connection as a  dual superfield variable, so that the dynamical
G-analyticity condition (or the Grassmann-harmonic zero-curvature
condition) for the first connection becomes a new equation of motion. We
shall use below the similar change of basic HSS variables for the $N=3$
SYM-theory.

In the HSS-approach to the $D=4,\;N=3$ SYM-theory \cite{GIKOS}, the
$SU(3)/U(1)\times U(1)$ harmonics have been used for a covariant
reduction of the spinor coordinates and derivatives and for the off-shell
description of the theory in terms of corresponding G-analytic
superfields. Moreover, it was shown that the $N=3$ SYM-constraints in the
ordinary superspace \cite{So} can been transformed to the zero-curvature
equations for the analytic harmonic gauge connections. It has been found
in Ref.\cite{RS} that there exist some set of analytic connections on HSS
to which no solutions of equations in the ordinary superspace  correspond.
However, nobody has succeeded to find interesting solutions of equations
for the harmonic connections.

For completeness, we shall remark two things. First, the alternative
$SL(2,C)$-harmonic  interpretation of the $N=3$ SYM-equations and the
corresponding harmonic zero-curvature equation for two G-analytic
connections was  considered in Ref.\cite{DO1}, however, the problem of
solving the $SL(2,C)$-harmonic equations will not be discussed in this
paper. Second, the very useful light-cone gauge conditions in the
ordinary superspace have been considered for solutions of the 4D self-dual
SYM-equations \cite{DL} and for the $10D$ SYM-equations \cite{GS}.
We shall discuss the analogous light-cone gauge conditions in the
harmonic approach.

It should be underlined that  the $N=3$ harmonic equations of motion in
the original formulation contain three independent G-analytic connections,
so they are more complicated than the $SU(2)$-harmonic  equations using
the single analytic connection. However, if we use the non-analytic
superfield matrix $v$ ( bridge) as a basic superfield variable of the
$N=3$ SYM-equations then the  zero-curvature equations for the harmonic
connections are solved automatically provided the dynamical G-analyticity
conditions should be imposed. The equivalent derivation of the same
dynamical bridge equations can be also considered in the
SYM-representation with the flat harmonic derivatives.

It will be shown that the most simple gauge condition for the matrix $v$
simplifies drastically  all basic equations, although the Lorenz
invariance is broken down to the $SO(1,1)$ subgroup by this condition. A
crucial feature of this gauge is the nilpotency of the bridge matrix,
namely that $v^3=0$.

In Sec.\ref{B}, we discuss the superfield constraints of the $D=4,\;
N=3$ SYM-theory and the light-cone gauge conditions for gauge connections
which simplify the SYM-equations.

Sec. \ref{D} is devoted to the analysis of the SYM-constraints in
the $SU(3)/U(1)\times U(1)$ harmonic superspace using the light-cone
nilpotent gauge for the bridge $v$. It will be shown that
a partial Grassmann decomposition of the non-analytic superfield
matrix $v$ is determined by three G-analytic matrices denoted by
$b^1, \bar{b}_3$ and $d^1_3$. The basic  equations for the
Grassmann connections in the bridge representation are equivalent  to the
set of G-analytic nonlinear harmonic differential equations for these
matrices. The nonlinear terms in our G-analytic SYM-equations are
proportional to Grassmann coordinates and therefore nilpotent. Using this
nilpotency one can obtain  linear 2-nd order differential constraints for
all three basic matrices.

In Sec.\ref{H} we analyze the equations for the harmonic connections
and superfield strengthes in a manifestly analytic representation.
 The analytic harmonic connections are nilpotent and contain the fermionic
matrices $b^1$ and $\bar{b} _3$ only. We construct also the non-analytic
harmonic connections in this representation and  a family of
geometric objects depending on the single coefficient $b^1$ of the
$v$-decomposition.

In the Appendices the basic formulas of the $SU(3)/U(1)\times U(1)$
harmonic superspace are summarized and the iterative procedure for solving
the SYM harmonic differential equations is considered. The resulting
finite number of the linear iterative harmonic equations can be , in
principle, directly solved. Since the harmonic expansions of the
SYM-superfields become very short on-shell the infinite number of
auxiliary fields vanishes. The harmonic equations in our approach admit a
dimensional reduction, so the further simplifications appear.

Thus, we find the simplest gauge conditions for the $N=3$ SYM-equations
in the harmonic superspace and show that these equations can be
transformed to the set of linear solvable matrix differential equations.
It is evident that the supersymmetry is crucial in our approach, and we
expect that the supersymmetric (Grassmann) version of integrability
relations yields new classes of the regular and stable SYM-solutions. It
will be interesting to analyze the formal reduction of the scalar and
fermion degrees of freedom in the exact constructions of these
SYM-solutions in order to study solutions of the non-supersymmetric
Yang-Mills equations.

\setcounter{equation}0
\section{\lb{B} $D=4,~N=3$ SYM
 constraints in reduced-symmetry representation}

The  coordinates of the $D=4,~N=3$ superspace are
\be
z^M=(x^{\alpha\dot\alpha} ,\theta^\alpha_i ,\bar\theta^{i\dot\alpha} )~,
\lb{A2b}
\ee
where $\alpha,~\dot\alpha$ are the $SL(2,C)$ indices and $i=1, 2, 3$  are
indices of the fundamental representations of the  group $SU(3)$.

We shall study solutions of the  SYM-equations using the non-covariant
notation for these coordinates
\bea
&&\xp\equiv x^{1\dot{1}} =t+x^3~,\q \xm\equiv x^{2\dot{2}}=t-x^3~,\q
y\equiv x^{1\dot{2}}=x^1+ix^2~,
\nn\\
&&\bar{y}\equiv x^{2\dot{1}} =x^1-ix^2~,\q(\tpi,~\tmi)\equiv
\theta^\alpha_i~,\q (\btpi,~\btmi)\equiv \bar\theta^{i\dot\alpha}
~.\lb{A2}
\eea
suitable when the Lorenz symmetry is reduced to $SO(1,1)$.
The general $N=3$ superspace has the odd dimension (6,6) in this
notation.

The coordinates have  the following $SO(1,1)$ weights (helicities)
\be
w(\xp)=2~,\q w(\xm)=-2~,\q w(y)=w(\bar{y})=0~,\q w(\theta^\pm_i)=
w(\bar\theta^{i\pm})=\pm1~.
\lb{A3}
\ee
and the simple conjugation properties
\be
(\xp)^\dagger=\xp~,\q (\xm)^\dagger=\xm~,\q y^\dagger=\bar{y}~,\q
(\tpi)^\dagger=\btpi~,\qquad (\tmi)^\dagger=\btmi \lb{A4}
\ee

 For products of arbitrary differential operators $X, Y$ or superfields
$f$ it is convenient to use the following  rules of conjugations:
\be
(X Y)^\dagger=Y^\dagger X^\dagger~,\qquad \overline{X f}=
-(-1)^{p(X)p(f)}\bar{X}\bar{f}~,\lb{A5}
\ee
where  $p(X)$ and $p(f)$ are the $Z_2$-parities.

 The algebra of $D=4,~N=3$ spinor derivatives in the reduced-symmetry
representation can be written in the form
\bea
&&\{\Dpk ,\Dpl\}=0~,\qq\{\bDpk ,\bDpl\}=0~,\q
 \{\Dpk ,\bDpl\}=2i\delta^k_l\Pp~,\nn\\
&&\{\Dmk ,\Dml\}=0~,\qq\{\bDmk ,\bDml\}=0~,\q
 \{\Dmk ,\bDml\}=2i\delta^k_l\Pm~,\lb{A6}\\
&&
 \{\Dpk ,\bDml\}=2i\delta^k_l\partial_y~,\q
\{\Dmk ,\bDpl\}=2i\delta^k_l\bar\partial_y~,
\q\{\Dpk ,\Dml\}=\{\bDpk ,\bDml\}=0~.\nn
\eea
Recall that the last two relations can contain six central charges,
however, in what follows we shall consider the basic superspace  without
central charges.

Let us  define the  gauge connections $A(z)$ and the corresponding
covariant derivatives $\nabla$ in the $(4|6,6)$-dimensional superspace
\bea
&&\nabla^i_\pm=D^i_\pm + A^i_\pm~,\qq\bar{\nabla}_{i\pm}=\bar{D}_{i\pm} +
\bar{A}_{i\pm}~,\lb{A7}\\
&&\nabla_\pp =\Pp + A_\pp ~,\qq\nabla_= =\Pm + A_= ~,\q
\nabla_y=\partial_y+A_y~,\q \bar\nabla_y=\bar\partial_y+\bar{A}_y~,\nn
\eea
then the $D=4,~N=3$ SYM-constraints \cite{So}
have the following reduced-symmetry form:
\bea
&&\{\nabla^k_+,\nabla^l_+\}=0~,\q \{\bar{\nabla}_{k+},\bar{\nabla}_{l+}\}
=0~,
\q \{\nabla^k_+,\bar{\nabla}_{l+}\}=2i\delta^k_l\nabla_\pp~,\lb{A8}\\
&&\{\nabla^k_+,\nabla^l_-\}=\bar{W}^{kl}
~,\q\{\nabla^k_+,\bar{\nabla}_{l-}\}=2i\delta^k_l \nabla_y~,\lb{A9}\\
&&\{\nabla^k_-,\bar{\nabla}_{l+}\}=2i\delta^k_l\bar{\nabla}_y~,\q
\{\bar{\nabla}_{k+},\bar{\nabla}_{l-}\}=W_{kl}~,\lb{A10}\\
&&\{\nabla^k_-,\nabla^l_-\}=0~,\q \{\bar{\nabla}_{k-},\bar{\nabla}_{l-}\}
=0~,
\q \{\nabla^k_-,\bar{\nabla}_{l-}\}=2i\delta^k_l\nabla_=~,\lb{A11}
\eea
where $W_{kl}$ and $\bar{W}^{kl}$ are the gauge-covariant superfield
strengthes constructed from the gauge connections. In particular,
this reduced form of the 4D constraints is convenient for the study of
dimensional reduction.

The equations of motion for the superfield strengthes follow from the
Bianchi identities
\bea
&& \nabla^i_\pm \bar{W}^{kl} + \nabla^k_\pm \bar{W}^{il}=0~,\nn\\
&& \bar{\nabla}_{i\pm} \bar{W}^{kl}={1\over2}(\delta^k_i
\bar{\nabla}_{j\pm}\bar{W}^{jl}- \delta^l_i\bar{\nabla}_{j\pm}
\bar{W}^{jk})~.\lb{A12}
\eea
Superfields $W_{kl}$ satisfy the conjugated equations.

Let us analyze first Eqs.\p{A8} together  with the relations
\be
[\nabla^k_+,\nabla_\pp]=[\bar\nabla_{k+},\nabla_\pp]=0~.
\ee

These equations for the positive-helicity connections have only the
following pure gauge solutions
\be
(\nabla^k_+, \bar\nabla_{k+}, \nabla_\pp)=g^{-1}(D^k_+, \bar{D}_{k+},
\Pp) g~.
\ee
Thus, when $g=1$ the simplest light-cone gauge conditions  can be taken
in the form
\be
A^k_+=0~,\q \bar{A}_{k+}=0~,\q A_\pp =0~.\lb{A14}
\ee
 Note that we do not discuss here  the off-shell
light-cone gauge superfields \cite{BLN}.

The analogous gauge conditions were considered in Ref.\cite{DL} for the
self-dual $4D$ SYM-theory and in Ref.\cite{GS} for the 10D SYM equations.
It should be underlined that these superfield gauge conditions break the
Lorenz group, but does not break the residual  invariance with respect to
translations $P_\pp,~P_=,~P_y,~\bar{P}_y$ and  supersymmetry generators
$Q^k_\pm,~\bar{Q}_{k\pm}$. Of course, one can choose  additional
non-supersymmetric gauge conditions for the remaining connections.
The group parameters of the residual local gauge invariance satisfy the
conditions
\be
(\Dpk,~\bDpk,~\Pp)\tau_r(z)=0~.\lb{A14r}
\ee

 Analogously to  Refs.\cite{DL,GS}, we can parametrize
all $N=3$ connections with $w=-1$ by the following matrix
superpotential $f_=$:
\be
A^k_-=\Dpk f_=~,\q \bar{A}_{k-}=\bDpk  f_=\lb{A15}
\ee

The equation
\be
\{\nabla^k_+,\bar{\nabla}_{l-}\}-{1\over3}\delta^k_l
\{\nabla^i_+,\bar{\nabla}_{i-}\}= 0\lb{A16}
\ee
is equivalent to the linear constraint
\be
\left(\Dpk\bDpl-{1\over3}\delta^k_l\Dpi \bDpi\right)f_= =0~.
\lb{A17}
\ee

Equations \p{A11}   give  the nonlinear relations
for  superpotential $f_=$
\bea
&&F^l_k-{1\over3}\delta^l_k F^i_i=0~,\q F^l_k\equiv (\bDmk\Dpl+
\Dml\bDpk)f_= -\{\Dpl f_= ,\bDpk f_=\} ~,\lb{A18}\\
&&(\Dmk\Dpl +\Dml\Dpk)f_= +\{\Dpk f_=,\Dpl f_=\}=0~,\lb{A19}\\
&&(\bDmk\bDpl +\bDml\bDpk)f_= +\{\bDpk f_=,\bDpl f_=\}=0~.\lb{A20}
\eea

Underline that this system of the second-order equations is equivalent
to the following set of the first-order equations:
\bea
&& \Dmk f_= =\Dpk\Omega +{1\over2}[f_=,\Dpk f_=]~,\nn\\
&& \bDmk f_= =\bDpk\Omega +{1\over2}[f_=,\bDpk f_=]~,
 \lb{A22}
\eea
with a suitable auxiliary superfield matrix function $\Omega$.

The equations for  superpotential $f_=$ can be analyzed directly,
however,  we prefer to use a harmonic-superspace method instead.

\setcounter{equation}0
\section{\lb{D}Harmonic-superspace equations for the nilpotent
 bridge matrix}

The Lorenz-covariant  $SU(3)/U(1)\times U(1)$ harmonic superspace was
introduced in Ref.\cite{GIKOS} for the off-shell description of the
$ N=3$ SYM-theory. The dynamical SYM-equations in this approach were
transformed into the set of pure harmonic equations for  G-analytic
superfield prepotentials.

Now we shall study  the $SU(3)/U(1)\times U(1)$ harmonic superspace in
another ( reduced-symmetry) representation which allows us to consider
the non-covariant gauges and the dimensional reduction.

We use  the $SU(3)$-matrix harmonic variables $u^I_i, u_I^i$ and the
analytic coordinates $\zeta=(X^\pp,~X^=,~Y,~\bar{Y}~| \theta^\pm_2,~
 \theta^\pm_3,~\bar\theta^{1\pm},~\bar\theta^{2\pm})$ which describes
the analytic superspace $H(4,6|4,4)$ (see Appendix A.1 ).

It is crucial that we start from the specific gauge conditions \p{A14}
for the $N=3$ SYM-connections which break  $SL(2,C)$, but preserve
$SU(3)$. Consider the harmonic transform of the  covariant Grassmann
derivatives  via the projections on the $SU(3)$-harmonics. As result we
get so called harmonized Grassmann covariant derivatives
\bea
&&\nabla^I_+\equiv u_i^I D^i_+=D^I_+\;,\q
\bar\nabla_{I+}\equiv u^i_I\bar{D}_{i+}=\bar{D}_{I+}\;,\q
\{D^I_+,\bar{D}_{K+}\}=2i\delta^I_K\Pp\;,\lb{D1b}\\
&&\nabla^I_-\equiv u_i^I\nabla^i_-=D^I_- +\cA^I_-\;,\q
\bar\nabla_{I-}\equiv u^i_I\nabla_{i-}=\bar{D}_{I-} +\bcA_{I-}\;,
\lb{D1}
\eea
with the harmonized Grassmann connections  defined by
\be
\cA^I_-=u_i^I A^i_-~,\q \bcA_{I-}=u^i_I \bar{A}_{i-}~.\lb{cbcon}
\ee

The $SU(3)$-harmonic projections of  superfield constraints
(\ref{A9}-\ref{A11}) can be derived from the basic set of the
$N=3$ zero-curvature (or G-integrability) conditions for two harmonized
 Grassmann connections:
\bea
&&\Dpo\cA^1_-=\bDph\cA^1_-=\Dpo\bcA_{3-}=\bDph\bcA_{3-}=0\;,\lb{D2}\\
&&\Dmo\cA^1_- +(\cA^1_-)^2=0\;,\q \bDmh\bcA_{3-}+(\bcA_{3-})^2=0\;,\nn\\
&&\Dmo\bcA_{3-}+\bDmh\cA^1_- +\{\cA^1_-,\bcA_{3-}\}=0\;.\lb{D3}
    \eea
All projections of the SYM-equations can be obtained  acting by the
harmonic $SU(3)$ derivatives $D^I_K$ on this basic set of conditions.

The G-integrability equations  have a very simple general
solution, namely
\be
\cA^1_-(v)=e^{-v}\Dmo e^v\;,\q \bcA_{3-}(v)=e^{-v}\bDmh e^v\;,\lb{D4}
\ee
where {\it the bridge } $v$ is a superfield matrix satisfying the
light-cone analyticity condition
\be
(\Dpo,\;\bDph) v=0\;,\lb{D5}
\ee
which is compatible with  on-shell representation \p{D1b}. Thus,
$v$ does not depend on $\tpo$ and $\btph$ in  analytic coordinates
\p{F6}.

Consider the gauge transformations of  bridge $v$
\be
e^v\;\Rightarrow\;e^\lambda e^v e^{\tau_r}\;,\lb{D11}
\ee
where $\lambda\in H(4,6|4,4)$ is a G-analytic matrix parameter,
and  parameter $\tau_r$  \p{A14r} does not depend on harmonics.
Matrix $e^v$  describes a map  of  gauge superfields
$ A^k_\pm, \bar{A}_{k\pm}$ defined in the central basis (CB) to the those
in the analytic basis (AB). By definition, the parameters of gauge
transformations in CB are independent of harmonics, while the gauge
parameters $\lambda$ are G-analytic. It is important that the off-shell
$N=3$ HSS formalism uses the same G-analytic gauge parameters.

The dynamical SYM-equations for $v$ due to Eqs. \p{D4}
are reduced to the following harmonic differential conditions for
the basic Grassmann connections:
\be
(D^1_2,\;D^2_3,\;D^1_3)\left(\cA^1_-(v),\;\bcA_{3-}(v)\right)=0\;.
\lb{Hanal}
\ee
Note that these {\it H-analyticity} relations are trivial for Grassmann
connections in the CB-superfield representation \p{cbcon}, but they
become the nontrivial differential equations for the connections
in the bridge representation. Equations for $v$
(\ref{Hanal}) are completely equivalent to the G-integrabity equations
\p{D3}, and they can be treated as a new representation of the
SYM-equations.

It is not difficult to built all Grassmann CB-connections in  terms of
basic ones, for example,
\bea
&&\cA_-^2(v)=D^2_1\cA_-^1(v)\;,\q
\cA_-^3(v)=D^3_1\cA_-^1(v)\;,\nn\\
&&\bcA_{1-}(v)=-D^3_1\bcA_{3-}(v)\;,\q
\bcA_{2-}(v)=-D^3_2\bcA_{3-}(v)\;.\lb{rcb1}
\eea

The harmonic projections of the non-Abelian CB-superfield strengthes
can also be constructed in the similar way
\bea
&&u^i_2u^k_3W_{ik}\equiv \cW_{23}=\bDpt\bcA_{3-}(v)=-\bDph\bcA_{2-}(v)\;,
\nn\\
&&u_i^1u^2_k\bar{W}^{ik}\equiv\bcW^{12}=-\Dpt\cA_-^1(v)=\Dpo\cA_-^2(v)~.
\lb{rcb2}
\eea

 By construction, these  CB-superfield strengthes satisfy the following
non-Abelian G-analyticity relations:
\bea
&&(\Dpo,\;\bDpt,\;\bDph)\cW_{23}=0\;,\q
(\nabla^1_-,\;\bar{\nabla}_{2-},\;
\bar{\nabla}_{3-})\cW_{23}=0\;,\lb{Grcb1}\\
&&(\Dpo,\;\Dpt,\;\bDph)\bcW^{12}=0\;,\q (\nabla^1_-,\;\nabla^2_-,\;
\bar{\nabla}_{3-})\bcW^{12}=0\lb{Grcb2}
\eea
and the simple H-analyticity conditions
\be
(D^1_2,\;D^2_3,\;D^1_3)(\cW_{23},\;\bcW^{12})=0
\;.\lb{Hanal2}
\ee
The  G- and H-analyticity conditions for non-Abelian superfields
 $\cW_{23}$ and $\bcW^{12}$ generate the corresponding component
SYM-equations of motion\footnote{The analogous harmonic projections of
 the $D=4,\;N=3$  superfield strengthes have been considered in
Ref.\cite{AFSZ} for the Abelian case. The Abelian harmonic superfields
$\cW_{23}$ and $\bcW^{12}$ describe the ultrashort on-shell
representations of the $N=3$ superconformal group.}.

Now we shall determine the explicit form of  bridge $v$. Using
the off-shell $(4,4)$-analytic $\lambda$-transformations
\be
\delta v=\lambda +{1\over2}[\lambda,v]+\ldots
\ee
one can choose the following non-supersymmetric nilpotent gauge condition
for $v$
\bea
&&v=\tmo b^1+ \btmh \bar{b}_3 +\tmo\btmh d^1_3\;,\q v^2=\tmo\btmh
[\bar{b}_3,b^1]\;,\q v^3=0~,\lb{D12c}\\
&&e^{-v}= I - v+{1\over2}v^2=I-\tmo b^1- \btmh \bar{b}_3+\tmo\btmh
({1\over2}[\bar{b}_3,b^1]-d^1_3 ) \;,
\lb{D12b}
\eea
where  fermionic matrices $b^1, \bar{b}_3 $ and bosonic matrix
$d^1_3 $ are analytic functions of  coordinates $\zeta$ introduced
in \p{F6}.

Note that  nilpotent gauges for  harmonic bridges are possible in the
harmonic formalisms with  off-shell analytic gauge groups only
\cite{GIKOS,GIK1}. Our gauge for $v$ combines the action of the analytic
gauge group and the light-cone analyticity condition \p{D5}.

Next we  shall use the  harmonic tilde-conjugation  (\ref{conj1},
\ref{conj3}) to describe the reality conditions for the gauge superfields
in HSS. For instance, the Hermitian conjugation $\dagger$ of the
superfield matrices includes transposition and this conjugation.
The  conditions for bridge $v$ in the gauge group  $SU(n)$ are
\be
\T v=0\;,\qq v^\dagger=-v\;,\lb{D15}
\ee
so that  matrices  $b^1, \bar{b}_3 $ and $d^1_3 $ have
the following properties in the group $SU(n)$:
\bea
&& \T b^1=0\;,\q \T \bar{b}_3=0\;,\q \T d^1_3=0\;
,\lb{D13}\\
&&(b^1)^\dagger=\bar{b}_3\;,\q (d^1_3)^\dagger=-d^1_3
\;.\lb{D14}
\eea

The fermionic matrices  have  specific properties
of traces
\be
\T (b^1)^{2k}=0~,\q \T (b^1\bar{b}_3)=-\T(\bar{b}_3b^1)~.
\ee

It is useful to consider the explicit parametrization of Grassmann
connection $\cA^1_-(v)$ and $\bcA_{3-}(v)$ in terms of  basic analytic
matrices \p{D12c}
\bea
&&\cA^1_-(v)\equiv e^{-v}\Dmo e^v=b^1-\tmo (b^1)^2+\btmh f^1_3
+\tmo\btmh[b^1,f^1_3]\,, \lb{K9}\\
&&\bcA_{3-}\equiv e^{-v}\bDmh e^v=\bar{b}_3 +\tmo \bar{f}^1_3-
\btmh (\bar{b}_3)^2+\tmo\btmh[\bar{f}^1_3,\bar{b}_3]~,
\eea
where the following auxiliary superfields are introduced:
\be
f^1_3=d^1_3-{1\over2}\{b^1,\bar{b}_3\}~,\q\bar{f}^1_3=-d^1_3-{1\over2}
\{b^1,\bar{b}_3\}~.\lb{K9b}
\ee

Equations $(\Dot, \Dth) \cA^1_-(v)=0$ generate the following
independent relations for the (4,4)-analytic matrices:
\bea
&& \Dot b^1=-\tmt (b^1)^2\;,\lb{D17}\\
&&\Dth b^1=-\btmt f^1_3~,\lb{D18}\\
&&\Dot f^1_3=\tmt[f^1_3,b^1]~,\q \Dth f^1_3=0~.
\lb{D19}
\eea

Equations $(\Dot, \Dth) \bcA_{3-}(v)=0$ are equivalent to the
relations
\bea
&&\Dot \bar{b}_3=\tmt\bar{f}^1_3~,\q \Dth \bar{b}_3=\btmt (\bar{b}_3)^2~,
\lb{D17b}\\
&&\Dot\bar{f}^1_3=0~,\q \Dth\bar{f}^1_3=\btmt[\bar{b}_3,\bar{f}^1_3]
\lb{D19b}~.
\eea
In the case of the  gauge group $SU(n)$, the last  equations are not
independent, but can be obtained by conjugation from the equations
for $b^1$ and $f^1_3$.

It is useful to derive the following relations for  matrices $b^1$ and
$\bar{b}_3$ which do not contain  auxiliary matrices $d^1_3, f^1_3$
or $\bar{f}^1_3$:
\bea
&&\tmt\Dot(b^1,~\bar{b}_3)=0~,\q \btmt\Dth ( b^1,~\bar{b}_3)=0~,
\lb{D23}\\
&&\tmt\Dth b^1+\btmt\Dot \bar{b}_3=\tmt\btmt\{b^1,\bar{b}_3\}~.\lb{D24}
\eea

Solutions of the linear equations for  matrices $f^1_3$ and
$\bar{f}^1_3$ satisfy the subsidiary condition
\be
f^1_3 +\bar{f}^1_3=-\{b^1,\bar{b}_3\}~.
\ee
The equation for  independent matrix $d^1_3$ is more complicated
than Eq.\p{D19}
\be
 \Dot d^1_3={1\over2}\tmt\left([d^1_3,b^1]
+{1\over2}[(b^1)^2,\bar{b}_3]\right)~.
\lb{D25}
\ee

The nilpotency of the nonlinear parts in these equations yields
 the subsidiary linear conditions  for the coefficient functions
\be
\Dot\Dot (b^1, \bar{b}_3, d^1_3)=0\;,\q \Dth\Dth (b^1, \bar{b}_3, d^1_3)
=0\;.\lb{lin2}
\ee
Note that one can  also find the following additional relations:
\bea
&&\Dot (b^1)^2=(\Dth)^2(b^1)^2=0~,\\
&&(\Dot b^1)^2=\Dot b^1 \Dot \bar{b}_3=\Dot b^1 \Dot d^1_3=0~.
\eea

The  harmonic linear equations for  analytic superfields
$b^1, \bar{b}_3,  d^1_3$ have simple (short) solutions, i.e. the
solutions with the finite number of the harmonic on-shell field
components. This shortness is an important property of the SYM-solutions
in the harmonic approach.

Now we shall study the action of the non-analytic harmonic
derivatives on the basic matrices. First, let us consider the following
relation:
\be
\Dot (\Dto)^2\cA^1_-(v)=(\Dto)^2\Dot \cA^1_-(v)=0~.
\ee
Equation $(\Dto)^2 \cA^1_-(v)=0$ produces the non-analytic
equations for the analytic matrices
\bea
&&(\Dto)^2b^1=\tmo(\Dto)^2(b^1)^2=2\tmo(\Dto b^1)^2~,\lb{2non}\\
&&(\Dto)^2f^1_3=\tmo(\Dto)^2[b^1,f^1_3]~.
\eea
Using  relation $\Dto \bcA_{3-}(v)=0$ one can obtain the following
equations:
\be
\Dto \bar{b}_3=\tmo \left(\Dto d^1_3+{1\over2}\{\bar{b}_3,\Dto b^1\}
\right)~,\q \tmo\Dto \bar{b}_3=0~.\lb{3non}
\ee

It is important that all differential harmonic equations
(\ref{D17}-\ref{D25}) contain  nilpotent elements $\tmt$ or $\btmt$ in
the nonlinear parts, so the simplest iteration procedure for finding
their solutions can be obtained  via a partial Grassmann decomposition.
In Appendix A.2, we consider the iterative procedure of solving
the basic non-Abelian harmonic differential equations for the
(4,4) analytic matrices $b^1$ and $\bar{b}_3$ (\ref{D23}-\ref{D24})
using the partial decomposition in the Grassmann variables $\tmt, \tmh,
\btmo, \btmt$. The matrix (4,0) coefficients of this decomposition have
dimensions $-1/2 \ge l \ge -5/2$.  The first iterative (4,0) equations
$(l=-1/2)$ are linear and homogeneous. The next harmonic iterative
equations for the  (4,0) components with $l \le -1/2$ are resolved in
terms of functions of the highest dimensions or their harmonic derivatives
and contain also the nonlinear sources constructed from the solutions of
the previous iterative equations. Note that some (4,0) iterative
equations are pure algebraic relations which reduce the number of
independent functions. The harmonic differential equations for the
independent (4,0) functions can be, in principle, explicitly solved using
the corresponding superfield Green functions. Thus, the $SU(3)/U(1)\times
U(1)$ harmonic method transforms the $N=3$ superfield SYM-constraints
together with the simple gauge conditions (\ref{A14}) to the non-Abelian
harmonic SYM-equations (\ref{Hanal}) which are equivalent
to the finite   set of the iterative solvable linear  equations.

Let us consider now the inverse harmonic transform which determines
the  on-shell gauge superfields in the ordinary superspace
\bea
&& \cA^1_-(v)~\Rightarrow~A^i_-(v)=(u^i_1+u^i_2D^2_1+u^i_3D^3_1)
e^{-v}\Dmo e^v\;\nn\\
&&\bcA_{3-}(v)~\Rightarrow~\bar{A}_{i-}(v)=(u_i^3-u^1_iD^3_1-u^2_iD^3_2)
e^{-v}\bDmh e^v\;,
\lb{K8}
\eea
where  definitions \p{rcb1} and  relations \p{F3} are used.
By construction, these superfields satisfy
the $D=4,~N=3$ CB-constraints
(\ref{A9}-\ref{A11}) and the harmonic differential
conditions
\be
D^I_K\left(A^i_-(v),\bar{A}_{i-}(v)\right)=0~,
\ee
if  equations \p{Hanal} or the equivalent equations for
 bridge $v$ are fulfilled.

\setcounter{equation}0
\section{\lb{H}Analytic representation of solutions }

In order to understand more deeply the geometric structure of our
harmonic-superspace solutions it is useful to represent them in
the analytic basis. Remember that the following covariant Grassmann
derivatives are flat in the analytic representation of the gauge group
before the gauge fixing:
\be
e^v\nabla^1_\pm e^{-v}\equiv\hat\nabla^1_\pm=D^1_\pm\;,\q
e^v\bar{\nabla}_{3\pm} e^{-v}\equiv\hat{\bar{\nabla}}_{3\pm}=
\bar{D}_{3\pm}\;.
\lb{H0}
\ee

The harmonic transform of the covariant derivatives via  matrix $e^v$
\p{D4} determines in AB the on-shell harmonic connections as a function
of $v$
\bea
&&\nabla^I_K\equiv e^v D^I_K e^{-v}=D^I_K+V^I_K(v)~,\nn\\
&&V^I_K(v)=e^v(D^I_K e^{-v})\lb{D10}~.
\eea
Note that the harmonic connections in the bridge representations satisfy
automatically the harmonic zero-curvature equations, for instance,
\bea
&& \Dot V^2_3-\Dth V^1_2 +[V^1_2,V^2_3]-V^1_3=0\;,\nn\\
&&\Dot V^1_3-\Doh V^1_2 +[V^1_2,V^1_3]=0\;.\lb{hzcr}
\eea

In the off-shell $N=3$ formalism \cite{GIKOS}, the connections
$V^1_2, V^2_3$ and $V^1_3$ are G-analytic by construction, so the
harmonic-zero curvature equations are interpreted as the basic equations
of motion. We prefer the bridge representation \p{D10}, since it is
directly connected with the classical SYM-solutions.

It is evident that  basic equations \p{Hanal} are equivalent
to the following set of the dynamic G-analyticity relations
for the composed harmonic connections:
\be
 (\Dmo,~\bDmh) \left(V^1_2(v), V^2_3(v), V^1_3(v)\right)=0\;,\lb{D7}
\ee

The positive-helicity analyticity conditions
\be
 (\Dpo,~\bDph)( V^1_2, V^2_3, V^1_3)(v)=0\lb{D12}
\ee
are satisfied automatically for  bridge $v$ in  gauge  \p{D5}.

The analytic SYM-equations (\ref{D17}-\ref{D19b}) are equivalent to
the following relations for the nilpotent on-shell analytic connections:
\bea
&&e^v\Dot e^{-v}=\tmt b^1\equiv V^1_2\;,\qq (V^1_2)^2=0\;,\lb{dyn}\\
&&e^v\Dth e^{-v}=-\btmt \bar{b}_3\equiv V^2_3\;,\qq (V^2_3)^2=0\;,
\lb{prep2}
\eea
which can be also rewritten as the  harmonic differential
equations
\be
(\Dot,~\Dth)e^v=-(V^1_2,~V^2_3)e^v~.
\ee

The Grassmann connections $a^I_\pm$ and $\bar{a}_{K\pm}$ in AB can be
calculated    analogously to
 Ref.\cite{Zu2}. We get
\bea
&&\hat{\nabla}^I_\pm=[\nabla^I_1,D^1_\pm]=D^I_\pm+a^I_\pm~,\q I=2, 3\\
&&a^2_\pm=-D^1_\pm V^2_1~,\q a^3_\pm=-D^1_\pm V^3_1~,\\
&&\hat{\bar{\nabla}}_{K\pm}=-[\nabla^3_K,\bar{D}_{3\pm}]=\bar{D}_{K\pm}
+\bar{a}_{K\pm}~,
\q K=1, 2\\
&&\bar{a}_{2\pm}=\bar{D}_{3\pm} V^3_2
\;,\q \bar{a}_{1\pm}=\bar{D}_{3\pm} V^3_1\lb{H0b}
\eea
where $V^2_1, V^3_1$ and $V^3_2$ are the non-analytic harmonic connections
defined in \p{D10}.

The nilpotent on-shell analytic connections satisfy the subsidiary
conditions
\be
\Dot V^1_2=\tmt\Dot b^1=0\;,\q \Dth V^2_3=-\btmt\Dth \bar{b}_3=0~.
 \lb{H1}
\ee

By using Eqs.\p{3non}   harmonic connection $V^2_1$ can be
written in terms of  superfield $b^1$ alone
\be
e^v\Dto e^{-v}=V^2_1=-\tmo \Dto b^1~.\lb{nonan4}
\ee

Connection $V^3_2$  can be obtained by conjugation of $V^2_1$
\be
V^3_2=-(V_1^2)^\dagger=-\btmh \Dht \bar{b}_3\;.\lb{H4}
\ee

Both connections  satisfy  the partial G-analyticity conditions
\be
\bar{D}_{3\pm}V^2_1=0\;,\q D^1_\pm V^3_2=0 \lb{H2}~.
\ee

 The  3-rd harmonic analytic connection can be readily calculated
\be
V^1_3=\Dot V^2_3-\Dth V^1_2+[V^1_2, V^2_3]=\tmh b^1-\btmo \bar{b}_3
~,\lb{3con}
\ee
where Eq.\p{D24} is used.
One can check straightforwardly the relations
\be
(V^1_3)^3=0~,\q\Doh V^1_3=\tmh\btmo\{b^1,\bar{b}_3\}~,\q (\Doh)^2 V^1_3
=0~.
\ee

The harmonic equations
\bea
&&\Doh b^1=-\tmh(b^1)^2-\btmo f^1_3~,\\
&&\Doh \bar{b}_3=\btmo (\bar{b}_3)^2+\tmh\bar{f}^1_3~,\\
&&\tmt\Doh b^1+\btmo\Dot \bar{b}_3=-\tmt\tmh(b^1)^2+\tmt\btmo
\{b^1,\bar{b}_3\}~,\\
&& \tmh\Doh b^1+\btmo\Doh \bar{b}_3=\tmh\btmo\{b^1,\bar{b}_3\}~,\\
&&\Doh f^1_3=\tmh [f^1_3,b^1]~.
\lb{h6}
\eea
can be also derived directly from Eqs.(\ref{D17}-\ref{D19b}).

Finally, for the last non-analytic harmonic connection we get
\be
V^3_1=\Dht V^2_1-\Dto V^3_2+[V^3_2,V^2_1]\equiv e^v\Dho e^{-v}~.
\ee

It is convenient to calculate the  superfield strength in AB
\be
\bar{w}^{12}=-\Dpo\Dmo V^2_1
=-\Dpt b^1~.
\lb{H5}
\ee
where Eq.\p{nonan4} is used.

Stress that the single coefficient matrix $b^1$ generates the family
 of the AB-geometric objects: $V^1_2, V^2_1, a^2_\pm$ and $\bar{w}^{12}$.
The conjugated $\bar{b}_3$-family of superfields contains $V^2_3, V^3_2,
\bar{a}_{2\pm}$ and
\be
w_{23}=-\bDph\bDmh V^3_2=\bDpt \bar{b}_3 \;.
\lb{H7}
\ee

The AB-superfield strengthes are directly connected with the
gauge CB-superfield strengthes (\ref{rcb2}), e.g.
$w_{23}=e^v\cW_{23} e^{-v}$.
These superfields satisfy the (4,2)-dimensional G-analyticity
conditions, for instance,
\be
D^1_\pm\bar{w}^{12}=\bar{D}_{3\pm}\bar{w}^{12}=D^2_\pm\bar{w}^{12}+
[a^2_\pm,\bar{w}^{12}]=0
\ee
and the non-Abelian H-analyticity conditions
\be
(\nabla^1_2, \nabla^2_3, \nabla^1_3)\bar{w}^{12}=0~.
\ee
The (4,2)-analytic superspaces have been considered earlier in Refs.
\cite{GIO,AFSZ}.

It should be noted that  function $d^1_3$ (or $f^1_3$) is an auxiliary
quantity in the framework of the analytic basis, since all harmonic
and Grassmann connections and tensors in this basis can be expressed by
means of superfields $b^1$ and $\bar{b}_3$ only. Nevertheless, the
construction of $d^1_3$ is important for the transition to the central
basis.

\setcounter{equation}0
\section{Conclusions and discussion}

Our method combines the light-cone gauge conditions for the $N=3$ gauge
superfields with the $SU(3)$-harmonic superspace approach. We have
described the harmonic transform of the $N=3$ SYM-equations of motion
in the standard superspace into the Grassmann integrability conditions for
the harmonized connections. All Grassmann and harmonic connections have
been constructed via the nilpotent superfield bridge matrix.

The nilpotency of nonlinear terms in the basic harmonic differential
SYM-equations simplifies drastically the iterative procedure of solving
these equations. Using the partial decomposition of the basic
(4,4)-superfields in terms of the Grassmann coordinates of negative
helicities we have obtained the finite set of solvable linear
(4,0)-equations which can be used for the explicit construction of
SYM-solutions.

The dimensional reduction of our harmonic equations simplify
the construction the $D<4$ solutions of the supersymmetric models with
12 supercharges. Since the $N=2$ superfield SYM-equations of motion have
the similar G-analyticity representation \cite{Zu4} the existence of
regular solutions with eight supercharges is not excluded.  It will be
interesting to analyze the formal reduction of the scalar and fermion
degrees of freedom in the exact constructions of  supersymmetric solutions
in order to estimate possibilities of our supersymmetric methods in
solving the non-supersymmetric Yang-Mills equations.

\vspace{1cm}

{\large \bf Acknowledgments}\\

The authors are grateful to E.A. Ivanov and E.S. Sokatchev for the
discussions.

This work is supported by the Votruba-Blokhintsev programme in Joint
Institute for Nuclear Research and it is also partially supported
by the grants GAASCR-A1010711, RFBR-99-02-18417,
RFBR-DFG-99-02-04022 and NATO-PST.CLG-974874.

\appendix
 \def\theequation{A.\arabic{equation}}
\setcounter{equation}0
\section{Appendices}
\subsection{\lb{F}  Reduced-symmetry form of $N=3$
 harmonic superspace}

The $SU(3)/U(1)\times U(1)$ harmonics \cite{GIKOS,GIO} parametrize
the corresponding 6-dimensional coset space. They form an $SU(3)$ matrix
$u^I_i$ and are defined modulo $U(1)\times U(1)$
\be
u^1_i=u^{(1,0)}_i\;,\q u^2_i=u^{(-1,1)}_i\;,\q
u^3_i=u^{(0,-1)}_i\;,\lb{F1}
\ee
where $i$ is the index of the triplet representation of $SU(3)$,
and the upper indices $I=1, 2, 3$ correspond to different combinations
of the $U(1)$-charges.

The complex conjugated harmonics $u^i_I$ have opposite $U(1)$ charges
and the upper index $i$ of the anti-triplet representation of $SU(3)$
\be
u^i_1=u^{i(-1,0)}~,\q u^i_2=u^{i(1,-1)}\;,\q u^i_3=u^{i(0,1)}\;.
\lb{F2}
\ee

These harmonics satisfy the following relations:
\bea
&& u_i^I u^i_J=\delta^I_J\;,\q u^I_i u^k_I=\delta^k_i\;,\nn\\
&&\varepsilon^{ikl}u_i^1 u_k^2 u_l^3=1\;.\lb{F3}
\eea

The $SU(3)$-invariant harmonic derivatives act on the harmonics
\bea
&&\partial^I_J u^K_i =\delta^K_J u^I_i\;,\q \partial^I_J u^i_K=
-\delta^I_K u^i_J\;,\lb{F4}\\
&&[\partial^I_J,\partial^K_L]=\delta^K_J\partial^I_L-\delta^I_L
\partial^K_J\;.\lb{F4b}
\eea

The operators of the $U(1)$ charges on the harmonics are
\be
h=\partial^1_1-\partial^2_2\;,\q h^\prime=\partial^2_2-\partial^3_3\;.
\lb{F4c}
\ee

One can consider the triplet of harmonic derivatives which annihilate
two harmonics
\be
(\partial^1_2,\partial^1_3,\partial^2_3)(u^1_i, u^i_3)=0\;.\lb{F5}
\ee
These harmonics are connected with the G-analyticity conditions.

We can define the real analytic harmonic superspace $H(4,6|4,4)$
with 6 coset harmonic dimensions $u^I_i$ and the following set of  4
even and (4+4) odd coordinates:
\bea
&&\zeta=(X^\pp,~X^=,~Y,~\bar{Y}~| \theta^\pm_2\;,
\theta^\pm_3\;,\bar\theta^{1\pm}\;,\bar\theta^{2\pm})\;,\qq
X^\pp=\xp +i(\tph\btph -\tpo\btpo)\;,\nn\\
&&X^= =\xm +
i(\tmh\btmh -\tmo\btmo)\;,\q Y=y+i(\tph\btmh -\tpo\btmo)\;,\nn\\
&&\bar{Y}=\bar{y}+i(\tmh\btph -\tmo\btpo)\;,\q
\theta^\pm_I=\theta^\pm_k u^k_I~,\q\bar\theta^{I\pm}=
\bar\theta^{\pm k}u_k^I~.
\lb{F6}
\eea
This superspace is covariant with respect to the $N=3$ supersymmetry
transformations.

The special $SU(3)$-covariant tilde-conjugation of harmonics
\be
u^1_i~\leftrightarrow~u^i_3\;,\q u^3_i~\leftrightarrow~u_1^i\;,\q
u^2_i~\leftrightarrow~-u_2^i\lb{conj1}
\ee
is compatible with  conditions  \p{F3}. On the harmonic derivatives
of an arbitrary harmonic function $f(u)$ this conjugation acts as follows
\be
\widetilde{\partial^1_3 f}=-\partial^1_3\widetilde{f}\;,\q
\widetilde{\partial^1_2 f}=\partial^2_3\widetilde{f}\;.
\lb{conj2}
\ee

The tilde-conjugation of the odd analytic coordinates has the following
form:
\be
\theta^\pm_1~\leftrightarrow~\bar\theta^{3\pm}~,\q
\theta^\pm_3~\leftrightarrow~\bar\theta^{1\pm}~,\q
\theta^\pm_2~\leftrightarrow~-\bar\theta^{2\pm}~.\lb{conj3}
\ee
Coordinates $X^\pp$ and $ X^=$ are real and $\widetilde{Y}=
\bar{Y}$.

The corresponding CR-structure involves the derivatives
\be
D^1_\pm,\;\bar{D}_{3\pm},\;\Dot,\;\Dth,\;\Doh=[\Dot,\Dth]\;,
\lb{F7}
\ee
which have the following explicit form in these coordinates:
\bea
&&D^{(1,0)}_\pm\equiv D^1_\pm=\partial^1_\pm\equiv
\partial/\partial\theta^\pm_1\;\q
\bar{D}^{(0,1)}_\pm\equiv \bar{D}_{3\pm}=\partial_{3\pm}\equiv
\partial/\partial \bar\theta^{3\pm}\;,\lb{F8}\\
&&D^{(2,-1)}\equiv \Dot =\partial^1_2
+i\tpt\btpo\Pp+i\tpt\btmo\PY+i\tmt\btpo\bPY
+i\tmt\btmo\Pm\nn\\
&&-\tpt\ppo-\tmt\pmo+\btpo\bppt+\btmo\bpmt~,\nn\\
&&D^{(-1,2)}\equiv\Dth =\partial^2_3
+i\tph\btpt\Pp+i\tph\btmt\PY+i\tmh\btpt\bPY
+i\tmh\btmt\Pm\nn\\
&&-\tph\ppt-\tmh\pmt+\btpt\bpph+\btmt\bpmh\;,\nn\\
&&D^{(1,1)}\equiv\Doh =\partial^1_3
+2i\tph\btpo\Pp+2i\tph\btmo\PY+2i\tmh\btpo\bPY
+2i\tmh\btmo\Pm\nn\\
&&-\tph\ppo-\tmh\pmo+\btpo\bpph+\btmo\bpmh\;,\lb{F9}
\eea
where $\Pp =\partial/\partial X^\pp,~\Pm =\partial/
\partial X^=,~\PY=\partial/\partial Y$ and $\bPY=
\partial/\partial\bar{Y}$.

One can construct also all harmonic derivatives in these coordinates
\bea
&&D^{(-2,1)}\equiv \Dto =\partial_1^2
-i\tpo\btpt\Pp-i\tpo\btmt\PY-i\tmo\btpt\bPY
-i\tmo\btmt\Pm~,\\
&&-\tpo\ppt-\tmo\pmt+\btpt\bppo+\btmt\bpmo\nn\\
&&D^{(1,-2)}\equiv\Dht =\partial_2^3
-i\tpt\btph\Pp-i\tpt\btmh\PY-i\tmt\btph\bPY
-i\tmt\btmh\Pm\nn\\
&&-\tpt\pph-\tmt\pmh+\btph\bppt+\btmh\bpmt\;,\lb{F9e}\\
&&D^{(-1,-1)}\equiv\Dho =\partial_1^3
-2i\tpo\btph\Pp-2i\tpo\btmh\PY-2i\tmo\btph\bPY
-2i\tmo\btmh\Pm\nn\\
&&-\tpo\pph-\tmo\pmh+\btph\bppo+\btmh\bpmo~, \lb{F9d}\\
&&H=[\Dot,\Dto]=h-\tpo\ppo-\tmo\pmo+\tmt\ppt+\tmt\pmt+\btpo\bppo
+\btmo\bpmo\nn\\
&&-\btpt\bppt-\btmt\bpmt~,\\
&&H^\prime=[\Dth,\Dht]=h^\prime-\tpt\ppt-\tmt\pmt+\tmh\pph+\tmh\pmh+
\btpt\bppt+\btmt\bpmt
\nn\\
&&-\btph\bpph-\btmh\bpmh
\eea
and the additional Grassmann derivatives
\bea
&& D^2_+ =\partial^2_+ +i\bar\theta^{2+}\Pp+i\bar\theta^{2-}\PY~,
\q  D^2_- =\partial^2_- +i\bar\theta^{2+}\bPY+i\bar\theta^{2-}\Pm~,\\
&&
D^3_+ =\partial^3_+ +2i\bar\theta^{3+}\Pp+2i\bar\theta^{3-}\PY~,
\q D^3_- =\partial^3_- +2i\bar\theta^{3+}\bPY+2i\bar\theta^{3-}\Pm~,\\
&&
\bar{D}_{1+} =\bar\partial_{1+} +2i\theta^+_1\Pp+2i\theta^-_1\bPY~,
\q \bar{D}_{1-} =\bar\partial_{1-} +2i\theta^+_1\PY+2i\theta^-_1\Pm~,\\
&&
\bar{D}_{2+} =\bar\partial_{2+} +i\theta^+_2\Pp+i\theta^-_2\bPY~,
\q \bar{D}_{2-} =\bar\partial_{2-} +i\theta^+_2\PY+i\theta^-_2\Pm~.
\eea

The tilde-conjugation of  harmonic derivatives $\Dot, \Dth$ and $\Doh$
is defined by analogy with relations \p{conj2}, and conjugation of the
Grassmann derivatives has the following form:
\be
D^1_\pm F~\leftrightarrow~-(-1)^{p(F)}\bar{D}_{3\pm}\widetilde{F}~,\q
D^2_\pm F~\leftrightarrow~(-1)^{p(F)}\bar{D}_{2\pm}\widetilde{F}~,\q
D^3_\pm F~\leftrightarrow~-(-1)^{p(F)}\bar{D}_{1\pm}\widetilde{F}~,
\ee
where $F$ is some superfield.

\subsection{ Analysis of iterative HSS equations}
Let us analyze
 the   analytic equations of Sec. \ref{D} for the basic fermionic
(4,4) matrices $b^1$ and $c_3$
\bea
&&\Dot b^1=-\tmt (b^1)^2~,\q (\Dot)^2b^1=(\Dth)^2b^1=0~,\lb{i1}\\
&&\Dth \bar{b}_3=\btmt (\bar{b}_3)^2~,\q (\Dot)^2\bar{b}_3=(\Dth)^2
\bar{b}_3=0~,\lb{i0}\\
&&\tmt\Dth b^1+\btmt\Dot \bar{b}_3=\tmt\btmt\{b^1,\bar{b}_3\}~,\lb{i2}\\
&&\tmh\Doh b^1+\btmo\Doh \bar{b}_3=\tmh\btmo\{b^1,\bar{b}_3\}~,
\lb{i3}\\
&&\tmh \Dth b^1+\btmt\Doh \bar{b}_3=-\btmo\btmt(\bar{b}_3)^2+
\tmh\btmt\{b^1,\bar{b}_3\}~.
\lb{i4}
\eea
The nonlinear terms in these equations contain the negative-helicity
Grassmann coordinates
$\tmt, \tmh, \btmo$ and $\btmt$, so a partial decomposition in terms of
these coordinates is very useful for the iterative analysis of solutions.
Equations for  auxiliary matrix $d^1_3$ do not give additional
restrictions on $b^1$ and $\bar{b}_3$.

Consider first the decomposition of harmonic derivatives \p{F9} and
define the harmonic derivatives  on the (4,0) analytic functions
depending on the analytic Grassmann variables  $\tpt,\tph,\btpo$ and
$\btpt$
\bea
&&\hDot=\partial^1_2+i\tpt\btpo\Pp-\tpt\ppo+\btpo\bppt~,\nn\\
&&\hDth=\partial^2_3+i\tph\btpt\Pp-\tph\ppt+\btpt\bpph~,\nn\\
&& \hDoh=\partial^1_3+2i\tph\btpo\Pp
-\tph\ppo+\btpo\bpph~.
\lb{I1}
\eea

The (4,0) decomposition of the  (4,4) analytic matrix function has the
following form:
\bea
&&b^1=\beta^1+\tmt B^{12}+\tmh B^{13}+\btmo B^0+\btmt B^1_2
+\tmt\tmh \beta^0+ \tmt\btmo \beta^2+\tmh\btmt \beta^{13}_2\nn\\
&&+
\tmh\btmo \beta^3+\btmo\btmt \beta_2
+\tmt\btmt\eta^1+\tmt\tmh\btmo B^{23}
+\tmh\btmo\btmt  B_2^3+\tmt\tmh\btmt C^{13}\nn\\
&&+\tmt\btmo\btmt C^0+\tmt\tmh\btmo\btmt\eta^3~.\lb{I2}
\eea
The analogous decompositions can be written for $\bar{b}_3$ and $d^1_3$.

It is easily to show that a part of the (4,0) coefficients
can be constructed as the algebraic functions of the basic set of
independent (4,0) matrices.
\bea
&& B^1_2=-\hDot B^0-i\tpt\PY\beta^1~,\lb{I4}\\
&&B^{12}=\hDth B^{13}+i\btpt\bPY\beta^1~,\lb{I5}\\
&&\beta^2=\hDth \beta^3-i\btpt\bPY B^0~,\lb{I6}\\
&&\eta^1=-\hDot\beta^2+i\btpo\bPY B^0
+i\tpt\PY B^{12}
+[\beta^1,B^0]
~,\lb{I8}\\
&&\beta^{13}_2=-\hDot \beta^3+i\tpt\PY B^{13}~,\lb{I9}\\
&& C^{13}=-\hDot B^{23}-i\btpo\bPY \beta^3-
\{\beta^1,\beta^3\}-[B^{13},B^1_2]~,
\lb{I10}\\
&& C^0+\bar{C}^0
=\hDot \bar{B}^2_1+\hDth B^3_2+i\Pm(\bar{B}^0-B^0)
+i\btpt\bPY\beta_2-i\tph\PY\beta^3
-i\btpo\bPY\bar\beta_1\nn\\&&
-i\tpt\PY \bar\beta^2
-\{\bar\beta_1,\beta^1\}-\{\bar\beta_3,\beta^3\}+[B^0,\bar{B}^0]
+[\bar{B}_{13},B^{13}]
~.\lb{I11}
\eea

Thus, the independent (4,0) matrix functions are
\be
B^0, B^{13}, B^{23}, (C^0-\bar{C}^0), B^3_2, \beta^1, \beta_2, \beta^3,
\beta^0,\eta^3\lb{I12}~.
\ee

The (4,0) matrix of  dimension $l=-1/2$ satisfies the linear
equations
\be
\hDot\beta^1= \hDth \beta^1=0~.\lb{I13}
\ee

The equations for the $l=-1$ matrices $B^{13}$ and $B^0$ are
\bea
&& \hDot B^{13}=0~,\q(\hDth)^2 B^{13}=0~,\nn\\
&&\hDoh B^{13}=-2i\btpo\bPY\beta^1-(\beta^1)^2~,\lb{I15}\\
&&\hDth B^0=0~,\q(\hDot)^2 B^0=0~,\nn\\
&&\hDoh (B^0-\bar{B}^0)
=i\tph\PY\beta^1+i\btpo\bPY\bar\beta_3+
\{\beta^1,\bar\beta_3\} ~.\lb{I16}
\eea
The inhomogeneous linear equations for $B^{13}$ and $B^0$ contain
sources with  functions $\beta^1$ and $\bar{\beta}_3=(\beta^1)^\dagger$
calculated on the previous stage.

The iterative equations for the (4,0) matrices with $l < -1$ can be
analyzed analogously. Each independent iterative equation is manifestly
resolved in terms of the harmonic derivatives of the corresponding
function and the sources of these equations  can  be calculated on the
previous stage of iteration.

Thus, it can be shown easily that the basic  $N=3$ harmonic  equations
for the  (4,4) analytic  functions  with the nilpotent nonlinear
terms are equivalent to the finite number of the
linear iterative (4,0) equations which contain non-Abelian sources
constructed from the solutions of the previous step of iteration.

Note that all iterative equations are simplified essentially for
the two-dimensional solutions which do not depend on the variables
$Y$ and $\bar{Y}$.

\end{document}